# Understanding the reduction of the edge safety factor during hot VDEs and fast edge cooling events

F.J. Artola[1], K. Lackner[2], G.T.A. Huijsmans[3,4], M. Hoelzl[2], E. Nardon[3], and A. Loarte[1]

[1] *ITER Organization, Route de Vinon sur Verdon, 13067 St Paul Lez Durance Cedex, France*
[2] *Max Planck Institute for Plasmaphysics, Boltzmannstr. 2, 85748 Garching, Germany*
[3] *CEA, IRFM, F-13108 St. Paul-lez-Durance cedex, France*
[4] *Eindhoven University of Technology, 5612 AP Eindhoven, The Netherlands*

Email: javier.artola@iter.org

**Abstract**

In the present work a simple analytical approach is presented in order to clarify the physics behind the edge current density behaviour of a hot plasma entering in contact with a resistive conductor. As it has been observed in recent simulations [1], when a plasma enters in contact with a highly resistive wall, large current densities appear at the edge of the plasma. The model shows that this edge current originates from the plasma response, which attempts to conserve the poloidal magnetic flux ($\Psi$) when the outer current is being lost. The loss of outer current is caused by the high resistance of the outer current path compared to the plasma core resistance. The resistance of the outer path may be given by plasma contact with a very resistive structure or by a sudden decrease of the outer plasma temperature (e.g. due to a partial thermal quench or due to a cold front penetration caused by massive gas injection). For general plasma geometries and current density profiles the model shows that given a small change of minor radius ($\delta a$) the plasma current is conserved to first order ($\delta I_p = 0 + \mathcal{O}(\delta a^2)$). This conservation comes from the fact that total inductance remains constant ($\delta L = 0$) due to an exact compensation of the change of external inductance with the change of internal inductance ($\delta L_{\text{ext}} + \delta L_{\text{int}} = 0$). As the total current is conserved and the plasma volume is reduced, the edge safety factor drops according to $q_a \propto a^2/I_p$. Finally the consistency of the resulting analytical predictions is checked with the help of free-boundary MHD simulations.

## 1 Introduction

The onset of vertical displacement events (VDEs) or axisymmetric modes in tokamaks is a well known phenomenon that arises in elongated plasmas. Uncontrolled VDEs can reduce the lifetime of the in-vessel components due to the large thermal and electromagnetic loads that they are able to produce. The type of VDEs with highest loads are often referred to as "hot" VDEs, where the plasma enters in contact with the wall keeping its initial thermal and magnetic energy. During such events, the plasma can develop additional asymmetric MHD instabilities that lead to a further localization and intensification of the loads. A particular concern for large scale tokamaks such as ITER and DEMO is the rotation of the electromagnetic load asymmetry. This rotation could resonate with the mechanical frequencies of the vessel creating a significant amplification of the vessel forces [2, 3]. The destabilization of toroidally asymmetric modes is typically associated with the onset of external MHD modes, which are driven by large edge current densities and a low edge safety factor [4]. Experimental observations of hot VDEs indicate that the edge safety factor ($q_a$) drops as the plasma cross-section is reduced [5], which is thought to cause the onset of external MHD modes. The change in $q_a$ and in the equilibrium parameters can lead to a fast stochastization of the field line topology and to a resulting thermal quench, which is observed with different large scale MHD simulation codes [6, 7, 8]. Relevant references summarizing the theory of disruptions are [9, 10]. In the present work, a simple analytical approach is presented in order to clarify the origin of the reduction of $q_a$ for hot VDEs as well as for fast cooling events of the plasma edge.

The kink edge safety factor [11] of an elongated plasma is defined as

$$q_a = 2\pi \left(\frac{1+\kappa^2}{2}\right) \frac{B_\phi}{\mu_0 R_0} \frac{a^2}{I_p} \qquad (1)$$

where $\kappa$ is the plasma elongation, $I_p$ is the total plasma current, $a$ is the plasma minor radius and $B_\phi$ is the toroidal magnetic field at the plasma geometric center placed at major radius $R_0$. This relation shows that for fixed $B_\phi$, $\kappa$ and $R_0$ a small variation of $q_a$ is given by

$$\frac{\delta q_a}{q_a} = -2\frac{\delta a}{a} - \frac{\delta I_p}{I_p} \qquad (2)$$

where the change of minor radius is imposed as $-\delta a$, with $\delta a$ positive. In order to decrease the edge safety factor over time, the following condition must be satisfied

$$\frac{\delta I_p}{I_p} = -\frac{\partial I_p}{\partial a}\frac{\delta a}{I_p} > -2\frac{\delta a}{a} \qquad \rightarrow \qquad \frac{\partial I_p}{\partial a} < 2\frac{I_p}{a} \qquad (3)$$

Therefore in order to know if $q_a$ will be reduced after a change of plasma minor radius $\delta a$, a relation between $\delta I_p$ and $\delta a$ is needed. Note that we consider only hot plasmas with a resistive decay time (at fixed minor radius) much longer than the time scales of interest. In the case where the resistive decay of $I_p$ is comparable to the considered time-scales, $q_a$ could increase even when $\partial_a I_p = 0$. In section 2 an analytical relation between $\delta I_p$ and $\delta a$ is obtained for highly conducting plasmas including the effect of a nearby conducting wall. The found expression is verified with the use of free-boundary MHD simulations in section 3 and finally our conclusions are shown in section 4.

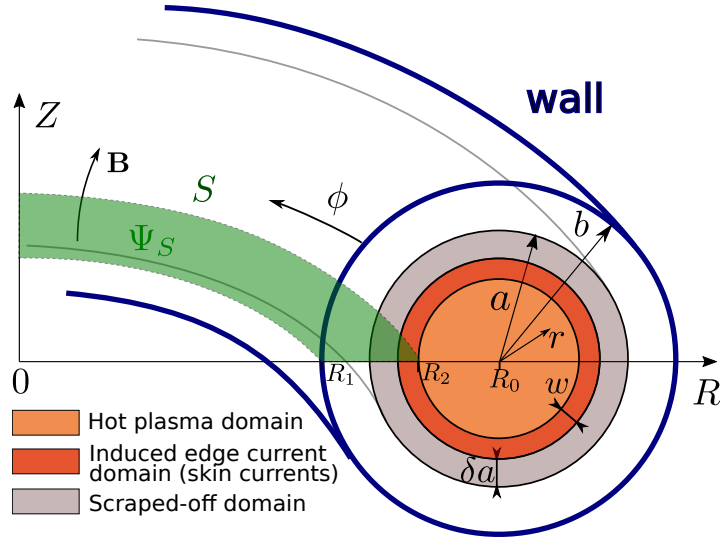

FIGURE 1: Geometry and coordinates for the analytical description of the scraping-off process.

## 2 Analytical calculation for $\delta I_p$

In the following we consider that a plasma has lost current in its outer region (it has been scraped-off) and that consequently its minor radius has been reduced by $\delta a$. This process could take place due to conduction of edge current through a highly resistive wall which has entered in contact with the plasma. This could be interpreted as a grounding effect. Equivalently a sudden drop of the edge plasma temperature due to a radiative collapse or due to the penetration of a cold pellet would cause a similar effect, which does not require a plasma motion into the wall. If the total resistance of the current path at the outer plasma is much larger than the core plasma resistance, the outer plasma current will vanish and the plasma minor radius will be reduced.

In order to describe the scraping-off process we consider the plasma defined by figure 1, which consists of three separated plasma regions

- **Scraped-off domain**: Existing from $r = a - \delta a$ to $r = a$. All plasma currents that were initially belonging to this region of thickness ($\delta a$) will vanish after the scraping-off process. Formally this is described by setting an infinite resistivity ($\eta = \infty$) at a certain time.
- **Hot plasma domain**: Existing from $r = 0$ to $r = a - (w + \delta a)$. This region is assumed to be ideally conducting on the considered time scales and therefore described by the setting ($\eta = 0$). It is also assumed that this region remains unperturbed and static during the scraping-off process.
- **Induced edge current domain**: Existing from $r = a - (w + \delta a)$ to $r = a - \delta a$. In this region of thickness ($w$) currents coming from the scraped-off domain can be induced. The width ($w$) can be approximated with the skin depth given by $w \sim \sqrt{\eta \tau / \pi \mu_0}$, where $\tau$ is the time-scale of the scraping-off mechanism. Therefore for ideally conducting plasmas ($\eta = 0$) or for very fast current losses ($\tau \to 0$) the skin depth becomes infinitely thin ($w \to 0$).

Note that figure 1 and the cylindrical radial coordinate $r$ are used for illustrative purposes, however the given geometry can be generalized to represent arbitrary shaped flux-surfaces ($\hat{r} = \hat{r}(r, \theta)$). Replacing $r$ by $\hat{r}$ gives identical final conclusions and expressions and therefore the symbol $r$ is used without loss of generality. In order to find the change of plasma current, we will use the conservation of poloidal magnetic flux that the hot plasma can provide. The time derivative of the poloidal magnetic flux ($\Psi_S$) contained in a toroidal ribbon ($S$) that extends from the major radius $R_1$ to $R_2$ is

$$\frac{\partial \Psi_S}{\partial t} = -2\pi \left[ R_2 E_\phi(R_2) - R_1 E_\phi(R_1) \right] \qquad (4)$$

where $E_\phi(R)$ is the toroidal electric field at the $R$ position. This equation can be found by applying Stoke's theorem to Faraday's law, where the flux is defined here as $\Psi = \int \mathbf{B} \cdot d\mathbf{S}$. In order to conserve the flux both terms on the RHS of equation (4) should vanish or cancel each other. The first term will vanish by choosing $R_2 = R_0 - (a - (\delta a + w))$ where the plasma is considered to be ideal (see figure 1) and therefore $E_\phi(R_2) = 0$. In ideal MHD, the electric field is zero in the frame of reference of the plasma ($\mathbf{E}' = \mathbf{E} + \mathbf{v} \times \mathbf{B} = 0$) even in the presence of flows. Consider a rigid displacement of the plasma column at a speed $\mathbf{v} = \mathbf{v}_{\text{VDE}}$. In this case, the flux conservation is still verified by placing ourselves into the frame of reference of the plasma. In this idealized model we only consider vertical rigid displacements and therefore the effects of equilibrium flows and complex VDE generated flows are not taken into account. By placing an ideally conducting wall at $R_1 = R_w^{in}$ the second term will also vanish as $E_\phi = 0$ inside ideal conductors. In the case without wall, the second term will vanish by extending the toroidal ribbon ($S$) to the origin of coordinates where $R_1 = 0$. Considering these choices for the boundaries of $S$ the flux $\Psi_S$ is conserved ($\partial_t \Psi_S = 0$).

Before the plasma current has been scraped-off, the poloidal flux within the toroidal ribbon $S$ can be decomposed as

$$\Psi_{S,0} = \Psi_{\text{bulk}} + \Psi_{\text{out}}^{\text{ext}} + \Psi_{\text{out}}^{\text{int}} \qquad (5)$$

where $\Psi_{\text{bulk}}$ is the flux produced by the hot plasma region and $\Psi_{\text{out}} = \Psi_{\text{out}}^{\text{ext}} + \Psi_{\text{out}}^{\text{int}}$ is the flux produced by the outer region, which includes both the region which will be scraped-off and the induced edge current domain (see figure 2 (a)). The total initial current within this outer region is represented by $\delta I_{\text{out}}$. After the reduction of the plasma minor radius, the final flux is

$$\Psi_{S,f} = \Psi_{\text{bulk}} + \Psi_{\text{skin}}^{\text{ext}} + \Psi_{\text{skin}}^{\text{int}} + \Psi_w \qquad (6)$$

where $\Psi_w$ is the flux produced by the ideally conducting wall ($\Psi_w = 0$ in case of no wall and $R_1 = 0$) and $\Psi_{\text{skin}} = \Psi_{\text{skin}}^{\text{ext}} + \Psi_{\text{skin}}^{\text{int}}$ is the flux produced by the induced edge current domain (see figure 2 (b)). By using the conservation of poloidal flux ($\Psi_{S,f} = \Psi_{S,0}$) we find

$$\Psi_{\text{skin}}^{\text{ext}} + \Psi_{\text{skin}}^{\text{int}} + \Psi_w = \Psi_{\text{out}}^{\text{ext}} + \Psi_{\text{out}}^{\text{int}} \qquad (7)$$

This expression can be used to calculate the final current in the induced edge current domain ($\delta I_{\text{skin}}$), which is related to the total change of plasma current by $\delta I_p = \delta I_{\text{skin}} - \delta I_{\text{out}}$. In order to do that we express the flux contributions as function of their associated currents

- $\Psi_{\text{out}}^{\text{ext}} = L_{\text{ext}}(a) \delta I_{\text{out}}$
- $\Psi_{\text{skin}}^{\text{ext}} = L_{\text{ext}}(a - \delta a) \delta I_{\text{skin}}$

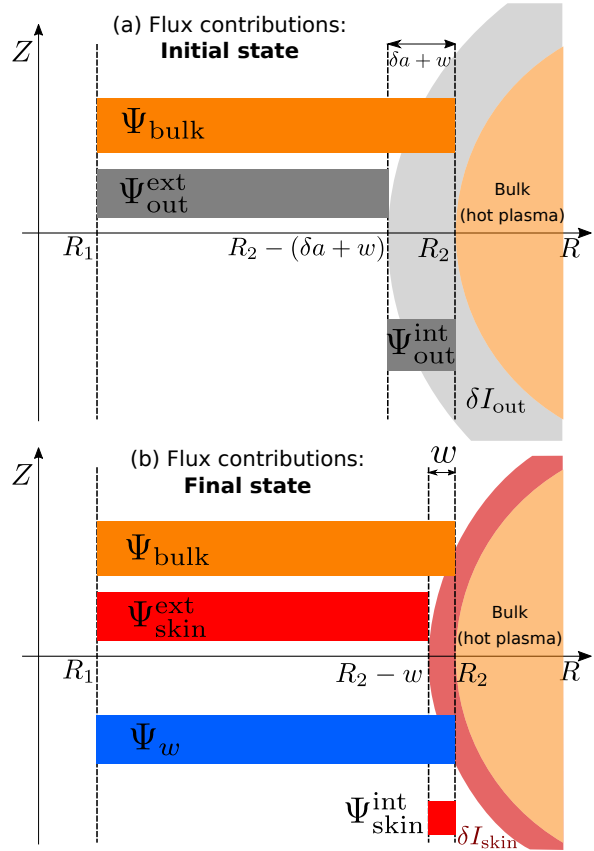

FIGURE 2: Flux contributions to $\Psi_S$ before (a) and after (b) the scraping-off of process. The colors associate the different current domains with their produced fluxes (coloured rectangles).

- $\Psi_{\text{out}}^{\text{int}} = c_{\text{out}} \delta I_{\text{out}} (\delta a + w)$
- $\Psi_{\text{skin}}^{\text{int}} = c_{\text{skin}} \delta I_{\text{skin}} w$
- $\Psi_w = -c_w \delta I_p$

where $L_{\text{ext}}(r)$ is the effective external inductance of a plasma with minor radius $r$ and $(c_{\text{out}}, c_{\text{skin}}, c_w)$ are geometrical constants. We define the effective plasma external inductance as $L_{\text{ext}}(r) = 2\pi \left[ \psi_p(R_0 - r) - \psi_p(R_1) \right] / I_p(r)$, where $\psi_p$ is the flux per radian produced by the plasma current $I_p(r)$. For the flux produced by the wall ($\Psi_w$) we have used that $\delta I_w \propto -\delta I_p$, which results from the fact that the magnetic field outside the wall is conserved due to the screening effect of the ideally conducting wall. With these considerations equation (7) is

$$(L_{\text{ext}}(a - \delta a) + c_{\text{skin}} w - c_w) \delta I_p = (L_{\text{ext}}(a) - L_{\text{ext}}(a - \delta a) + c_{\text{out}} \delta a + (c_{\text{out}} - c_{\text{skin}}) w) \delta I_{\text{out}} \quad (8)$$

and finally by regrouping terms and using the following Taylor expansion $L_{\text{ext}}(a) = L_{\text{ext}}(a - \delta a) + \partial_a L_{\text{ext}} \delta a$ we find

$$\delta I_p = \frac{(\partial_a L_{\text{ext}} + c_{\text{out}}) \delta a + (c_{\text{out}} - c_{\text{skin}}) w}{L_{\text{ext}}(a) - \partial_a L_{\text{ext}} \delta a + c_{\text{skin}} w - c_w} \delta I_{\text{out}} \quad (9)$$

Considering that $L_{\text{ext}}(a)$ is a quantity of zero order, for finite initial edge current densities $\delta I_{\text{out}} \propto J_{\text{edge}} \delta a$ and $w \sim \delta a$ we find that

$$\delta I_p = \mathcal{O}(\delta a^2) \quad (10)$$

which demonstrates that for ideally conducting plasmas with arbitrary shape the current is conserved up to second order. We define the quantity $\epsilon_{\text{ind}}$ as the relative amount of re-induced current as

$$\epsilon_{\text{ind}} \equiv \frac{\delta I_{\text{skin}}}{\delta I_{\text{out}}} = 1 + \frac{(\partial_a L_{\text{ext}} + c_{\text{out}}) \delta a + (c_{\text{out}} - c_{\text{skin}}) w}{L_{\text{ext}}(a) - \partial_a L_{\text{ext}} \delta a + c_{\text{skin}} w - c_w} \quad (11)$$

With this definition, $\epsilon_{\text{ind}} = 1$ if $\delta I_p = 0$ and all the current from the scraped-off region is re-induced into the new plasma boundary and $\epsilon_{\text{ind}} = 0$ when $\delta I_p = -\delta I_{\text{out}}$.

The conclusion drawn from equation (10) can be interpreted as a first order cancellation of the change of external inductance ($\delta L_\text{ext}$) with the change of internal inductance $\delta L_i$). As the total plasma flux ($\Psi$) contained in the toroidal ribbon that extends from $R_1$ to $R_0$ is conserved $\delta \Psi = \delta(LI_p) = 0$, the total current is conserved up to second order only if $\delta L = \mathcal{O}(\delta a^2)$, which implies that $\delta L_\text{ext} + \delta L_i = \mathcal{O}(\delta a^2)$.

Note however that the scraped-off current ($\delta I_\text{out}$) can grow to order zero as the edge current builds up from previous reductions of the minor radius. If $\delta I_\text{out} = \mathcal{O}(I_p)$, the change of total current grows to first order according to equation (9). In the case where ($w \ll \delta a$), the rate at which the current is lost with respect to the change of minor radius is

$$\frac{\partial I_p}{\partial a} \approx -\frac{\partial_a L_\text{ext} + c_\text{out}}{L_\text{ext}(a)} \delta I_\text{out} = -\frac{\partial_a L_\text{ext} + c_\text{out}}{L_\text{ext}(a)} (\Delta I_\text{out} + \Delta I_p) \tag{12}$$

where $\Delta I_p(a) \equiv I_p(a) - I_p(a_0)$ is the total change of plasma current and $\Delta I_\text{out} \equiv I_0(a_0) - I_0(a)$ is the total scraped-off current with respect to the initial plasma state with radius $a_0$ and current profile $I_0(r)$. Note that the non-indexed minor radius is a function of time $a(t)$. After a time interval $\delta t$, the new radius is $a(t+\delta t) = a(t) - \delta a$, with $\delta a$ being positive. The subscript "0" indicates quantities at the initial plasma before any plasma "peeling" has taken place $a_0 = a(t=0)$. Similarly the profile of total toroidal current as a function of the radial coordinate $I(r,t)$ at t=0 is $I_0(r)$. In the latter expression we have used that $\delta I_\text{out}$ is the current in the skin layer before the small $\delta a$ reduction ($\delta I_\text{out} = I_\text{skin}(t - \delta t) = \Delta I_\text{out} + \Delta I_p$).

## Limit for a circular plasma with large aspect ratio

For a large aspect ratio circular plasma enclosed by a wall with minor radius ($b$) the previous quantities are

$$\hat{L}_\text{ext}(a) \equiv \frac{L_\text{ext}(a)}{\mu_0 R_0} = \begin{cases} [\ln(8R_0/a) - 2] & \text{no wall} \\ \ln(b/a) & \text{ideal wall} \end{cases}$$

- $\partial_a L_\text{ext} = -\mu_0 R_0/a$
- $c_\text{out} = c_\text{skin} = \mu_0 R_0/(2a)$
- $c_w = 0$
- $\delta I_\text{out} = 2(J_\text{edge}/<J_0>)(\delta a + w)I_p/a$

where $<J_0>$ is the initial averaged current density $<J_0> \equiv I_p/(\pi a^2)$ and $J_\text{edge}$ is the initial current density in the scraped-off domain. For the calculation of ($c_\text{out}, c_\text{skin}$) we have assumed that the current density in the perturbed region is spatiality constant. The flux within the wall produced by the wall itself is taken to be zero ($c_w = 0$) as it is expected from a circular wall in the large aspect ratio assumption [12]. With these quantities equations (2), (9) and (11) give

$$\delta I_p = -\frac{J_\text{edge}/<J_0>}{\hat{L}_\text{ext}(a) + \delta a/a + w/(2a)} \frac{\delta a}{a} \frac{\delta a + w}{a} I_p \tag{13}$$

$$\epsilon_\text{ind} = 1 - \frac{1}{2} \frac{\delta a/a}{\hat{L}_\text{ext}(a) + \delta a/a + w/(2a)} \tag{14}$$

$$\frac{\delta q_a}{q_a} = \left(-2 + \frac{J_\text{edge}/<J_0>}{\hat{L}_\text{ext}(a) + \delta a/a + w/(2a)} \frac{\delta a + w}{a}\right) \frac{\delta a}{a} \tag{15}$$

and from equation (13), the rate at which $I_p$ varies with $a$ is

$$\frac{\partial I_p}{\partial a} = \frac{1}{2a} \frac{1}{\hat{L}_\text{ext}(a) + \delta a/a + w/(2a)} \delta I_\text{out} \tag{16}$$

For finite $\hat{L}_\text{ext}(a)$, the latter expression together with (3) gives a condition for the minimum value of $\delta I_\text{out}$ to stop the drop of $q_a$

$$\frac{\delta I_\text{out}}{I_p} > 4\hat{L}_\text{ext}(a) \tag{17}$$

For a typical tokamak aspect ratio of 3, the latter expression implies that the accumulated edge current should be 4.7 times larger than the total plasma current, which is an absurd situation that proves that the safety factor always decreases.

Nevertheless the effect of a close enough ideally conducting wall can change this situation as $\hat{L}_{\text{ext}}(a) \to 0$ when $b/a \to 1$. If the ideal wall is located exactly at the plasma boundary ($b/a = 1$) and $w = 0$, half of the current in the scraped-off region can be re-induced in the wall ($\epsilon_{\text{ind}} = 1/2$). For this special case the change of the edge safety factor is

$$\frac{\delta q_a}{q_a} = \left(-2 + \frac{J_{\text{edge}}}{<J_0>}\right) \frac{\delta a}{a} \qquad (18)$$

which gives $\delta q_a/q_a > 0$ if $\frac{J_{\text{edge}}}{<J_0>} > 2$.

## 3 Simulations

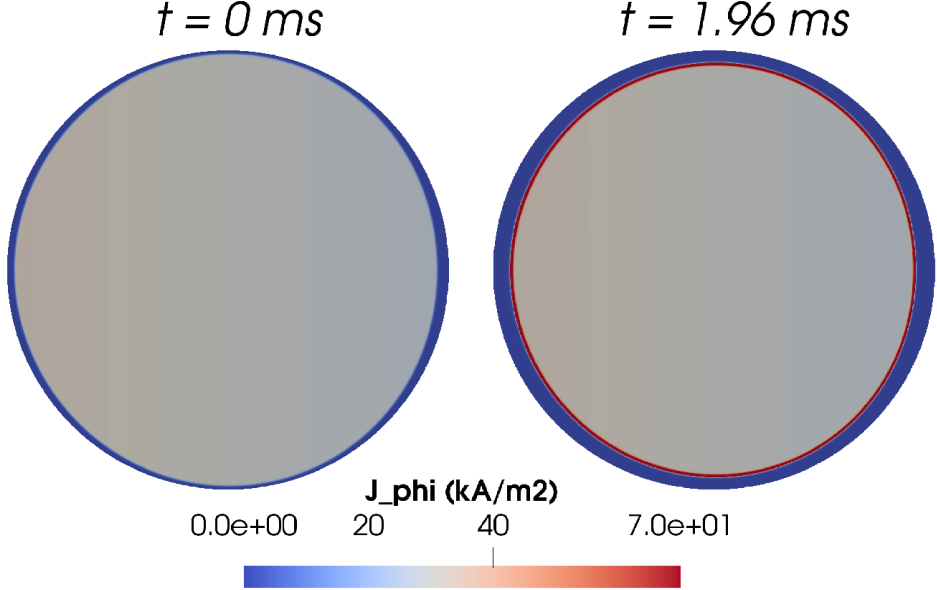

FIGURE 3: Current density contour plots for the case without wall at $t = 0$ ms and at $t = 1.96$ ms.

In order to check the consistency of the previous analytical expressions we perform simulations with the reduced MHD model of the non-linear code JOREK [13]. A circular plasma is constructed with an initially flat current density profile (see figure 3), a minor radius ($a = 0.97$ m), a major radius ($R_0 = 10$ m) and a poloidal beta ($\beta_p = 0.004$). The plasma is held in toroidal equilibrium thanks to the vertical field produced by a set of external PF coils situated at $R = 8$ m and at $R = 12$ m. Free-boundary conditions are used by coupling JOREK with the code STARWALL [14], which allow the plasma to move radially and vertically. The boundary condition for the toroidal current density is simply given by Ampère's law ($J_\phi = -\Delta^*\psi/(\mu_0 R)$). The chosen polar grid is formed by 140 (radial) x 42 (poloidal) Bézier elements with radial refinement in order to resolve the thin skin currents that appear at the edge of the plasma. The resistivity is chosen to be independent of the plasma temperature and it is given by

$$\eta = \eta_0 + \eta_Z/2 \left(1 + \tanh\left[(Z_{\text{th}} - Z)/\sigma_Z\right]\right) \qquad (19)$$

where $\eta_0 = 1.95 \times 10^{-8} \Omega$ m, $\eta_Z = 10^4 \eta_0$, $Z_{\text{th}} = -0.95$ m and $\sigma_Z = 0.003$ m. The second term is chosen to act as the scraping-off mechanism, which creates a highly resistive region at the lower plasma edge that makes the current decay in a fast time-scale compared to the resistive time of the plasma core. With this setting, the resulting change of minor radius is $\delta a = 3.5$ cm. As it can be seen in figure 3, after 1.96 ms, skin currents appear at the edge of the plasma. The resulting thickness of the skin current during this time is $w = 2.3$ cm.

An ideal wall is placed near the plasma at different minor radii ($b$). The obtained $\delta I_p/I_{p0}$ and $\epsilon_{\text{ind}}$ in the simulations are shown together with the analytical prediction given by equations (13) and (14) in figure 4. The change of safety factor is shown in figure 5 where the analytical prediction is given by equation (15). The results show an excellent agreement between theory and simulations which validates the analytical model.

More complex JOREK simulations also show this behaviour as it can be observed in the following VDE simulation where a hot NSTX plasma enters in contact with a highly resistive wall (see figure 6), which has also been used in benchmarks with other non-linear MHD codes recently in [15]. During the simulation, a Spitzer-like resistivity profile $\eta = \eta_0(T_{e0})(T_{e0}/T_e)^{3/2}$ is chosen where the core electron temperature is $T_{e0} = 1$ keV. The temperature at the wall has a fixed value of $T_e = 1$ eV, which imposes a large plasma resistivity at the wall that scrapes-off the plasma edge when the plasma becomes limited. For the chosen wall resistivity, the plasma minor radius is reduced by a factor 2 in 6 ms, which is a short enough time-scale to consider the plasma as highly conducting for the given plasma resistivity.

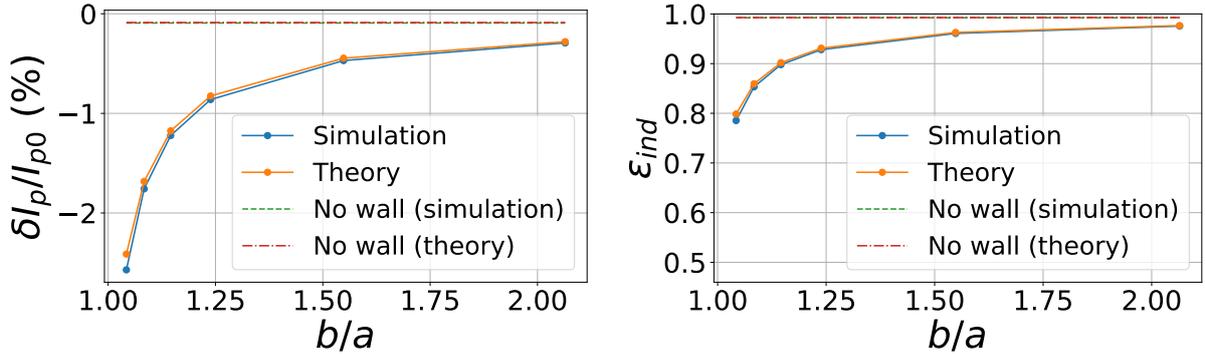

Figure 4: Simulation results and analytical predictions for $\delta I_p/I_{p0}$ (left) and $\epsilon_{\text{ind}}$ (right) for different ideal wall minor radii.

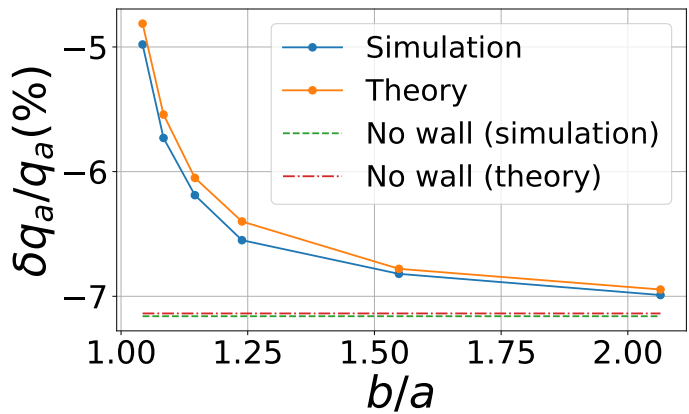

Figure 5: Simulation results and analytical predictions for $\delta q_a/q_a$ for different ideal wall minor radii.

As it can be inferred from figure 7 (b), the evolution of the safety factor is dominated by the change of plasma minor radius and the change of $I_p$ starts influencing $q_a$ when the minor radius has already dropped by a factor 2. Here the averaged minor radius is defined as $<a> \equiv l_p/(2\pi)$, where $l_p$ is the length of the poloidal flux contour of the plasma boundary. In figure 7 (a) the obtained $\partial I_p/\partial a$ is compared with the prediction given by equation (16), where the effective external inductance is calculated from the simulation with $L_{\text{ext}}(a) = 2\pi\psi_{p,\text{bnd}}/I_p$ as well as $\delta I_{\text{out}}$. Although the model and the simulation results do not agree as well as for the simple case shown in figure 3, the analytical model gives the correct behaviour and order of magnitude of $\partial I_p/\partial a$. Discrepancies may be caused by the effect of a finite wall conductivity or the effect complex VDE generated flows breaking the rigid displacement assumption. The assessment of these effects is left for future work.

## 4 Conclusions

In the typical case where the effective external inductance $L_{\text{ext}}(a)$ and the edge current density $J_{\text{edge}}$ are finite, we have demonstrated that $\delta I_p = \mathcal{O}(\delta a^2)$ for highly conducting plasmas that are scraped-off by a change of plasma radius of $\delta a$. In this case the edge safety drops according to

$$\frac{\delta q_a}{q_a} = -2\frac{|\delta a|}{a} + \mathcal{O}(\delta a^2/a^2) < 0 \qquad (20)$$

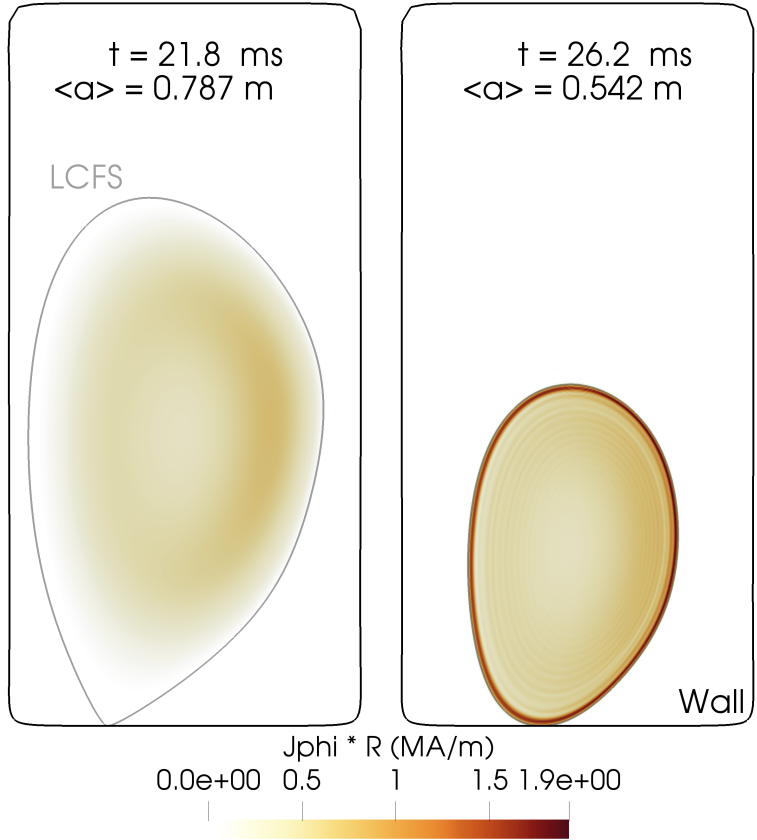

FIGURE 6: NSTX hot VDE at two different times showing the induction of edge currents. The toroidal current density is shown together with the wall boundary (black line) and the last closed flux surface (grey line).

The found analytical expression for $\delta I_p$ for general plasmas (equation (9)) has been validated in the limit of a circular plasma with large aspect ratio with the help of free-boundary MHD simulations. Additional simulations for a realistic NSTX plasma qualitatively agree with the analytical predictions and prove the validity of (20). The change of current ($\delta I_p$) can grow to first order in $\delta a$ when the accumulated edge current is comparable to the total current $\delta I_{\text{out}} = \mathcal{O}(I_p)$. Nevertheless for typical tokamak values of $L_{\text{ext}}(a)$, the decay of $I_p$ cannot stop the reduction of the edge safety factor (equation (17)). The model and the simulations show that the plasma current is reduced by 15% after decreasing its minor radius by a factor 2, which for typical initial edge safety factors ($q_a \sim 3$) implies that $q_a < 1$ before $a < a_0/2$.

In the special case where a highly conducting wall is placed at the plasma boundary ($L_{\text{ext}}(a) \to 0$), half of the plasma current in the scraped-off domain can be induced in the wall implying that $\delta I_p = \mathcal{O}(\delta a)$. In this situation $\delta q_a$ is given by equation (18). Note however that during "hot" VDEs the surrounding conducting structures are highly resistive compared to the plasma by definition. Therefore during "hot" VDEs, $q_a$ should evolve according to equation (20). Nevertheless the ideal wall effect could play a role during phenomena like fast cold front penetration given by massive gas injection (MGI) or by shattered pellet injection (SPI) as planned for mitigation of such disruption events [16].

## Acknowledgements

This work was supported by the ITER Monaco Fellowship. ITER is the Nuclear Facility INB no. 174. This paper explores physics processes during the plasma operation of the tokamak when disruptions take place; nevertheless the nuclear operator is not constrained by the results presented here. The views and opinions expressed herein do not necessarily reflect those of the ITER Organization.

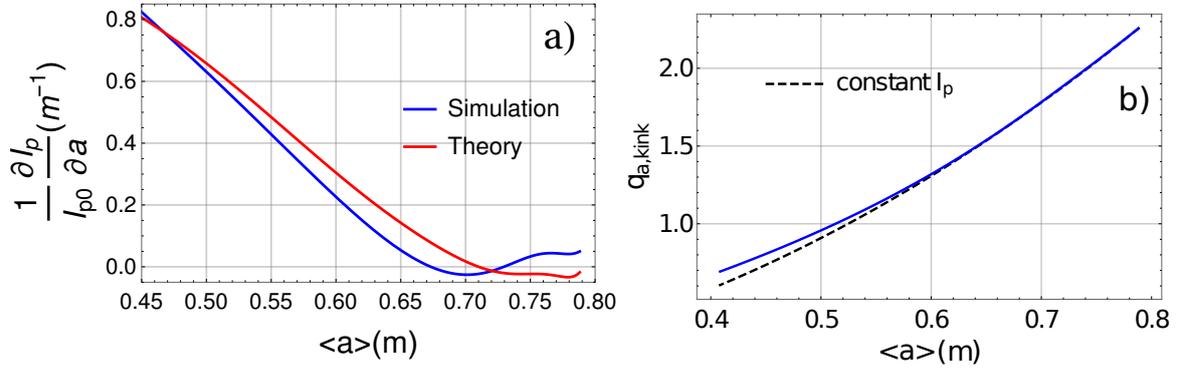

FIGURE 7: (a) Rate of change of plasma current with respect to the change of minor radius and (b) computed kink edge safety factor as a function of the averaged minor radius during an NSTX VDE simulation. The resulting $q_a$ computed with (1) is compared with its value assuming a constant $I_p$ during the simulation (dashed curve).